\documentclass{IEEEtran}
\usepackage{cite}
\usepackage{amsmath,amssymb,amsfonts}
\usepackage{algorithmic}
\usepackage{graphicx}
\usepackage{tabularx}
\usepackage{bbding}
\usepackage{multirow}
\usepackage{booktabs}
\usepackage[squaren]{SIunits}
\usepackage{longtable}
\usepackage{rotating}
\usepackage{makecell}
\usepackage{pdflscape}
\usepackage{adjustbox}
\usepackage{ulem}
\usepackage{bm,array}
\usepackage[dvipsnames]{xcolor}
\usepackage{caption}
\usepackage{subcaption}
\usepackage{color,soul}
\usepackage{hyperref}
\usepackage[roman]{parnotes}
\makeatletter
\def\parnoteclear{%
    \gdef\PN@text{}%
    \parnotereset
}
\makeatother

\begin{document}

\title{An UAV-based Experimental Setup for Propagation Characterization in Urban Environment}

\renewcommand\footnotemark{}
\renewcommand\footnoterule{}

\author{
Franco Fuschini,    
Marina Barbiroli, 
Enrico M. Vitucci,~\IEEEmembership{~Senior~Member,~IEEE}, 
Vasilii Semkin, 
\\Claude Oestges,~\IEEEmembership{Fellow,~IEEE},
Bruno Strano,
and Vittorio Degli-Esposti,~\IEEEmembership{Senior~Member,~IEEE}      
}                                     

\thanks{Manuscript received January XX, 2020; revised {\ldots}.} \thanks{(\textit{Corresponding author: Franco Fuschini})}
\thanks{The work of V. Semkin was partly supported in part by the Academy of Finland and the Jorma Ollila grant.}
\thanks{F.	Fuschini, M. Barbiroli, E. M. Vitucci, B. Strano and V. Degli-Esposti are with the Dept. of Electrical, Electronic and Information Engineering "G. Marconi", University of Bologna, IT-40136 Bologna, Italy (e-mail: franco.fuschini, marina.barbiroli, enricomaria.vitucci, bruno.strano, v.degliesposti\}@unibo.it}
\thanks{V.	Semkin is with the VTT Technical Research Centre of Finland Ltd, 02150 Espoo, Finland (e-mail: vasilii.semkin@vtt.fi).}
\thanks{Claude Oestges is with the Universit\`{e} catholique de Louvain, B-1348 Louvain-la-Neuve, Belgium (e-mail: claude.oestges@uclouvain.be).}

\maketitle

\begin{abstract}
A measurement setup including millimeter-wave and ultra wideband transceivers mounted on both a customized UAV and a ground station and a measurement procedure for full 3D wireless propagation analysis is described in this work. The custom-made developed system represents a flexible solution for the characterization of wireless channels and especially of urban propagation, as the drone might be easily located almost anywhere from ground level to the buildings rooftop and beyond. The double directional properties of the channel can be achieved by rotating directive antennas at the link ends. Other possible applications in urban contexts include the study of outdoor-to-indoor penetration, line-of-sight to non-line-of-sight transition, 3D scattering from buildings and air-to-ground channel characterization for UAV-assisted wireless communications.
\end{abstract}


\maketitle

\section{Introduction}
\label{sec:introduction}
Next-generation wireless systems will be deployed in a large variety of propagation environments (from indoor pico-cells to large cells served by airborne base stations), over multiple frequency bands (from UHF to THz frequencies) and with different antenna configurations (from small antennas for IoT devices to massive arrays). Since the radio channel characteristics can vary very much over such a wide range of different scenarios, a large number of measurements in very different spatial configurations is necessary to characterize the radio channel for the design and the deployment of future wireless networks. Traditionally, propagation measurements have been carried out moving the measurement setup terminal around the environment using vehicles or trolleys. However, the use of Unmanned Aerial Vehicles (UAV) to position the radio terminals in an arbitrary location in a 3D space and to steer directive antennas in different directions virtually without limitations is an attractive solution, especially for the characterization of urban radio channels. As a matter of fact, the drone can mimic a base station or an user equipment that might be located anywhere from ground level to the top of the highest buildings or at mid-air on a traffic light between buildings, and allows for the maximum positioning flexibility.

Besides being useful to carry out 3D propagation measurements, the use of low-altitude UAV has been considered in many different applications, like public safety surveillance, search and rescue operations \cite{ref1}, radio frequency sensing and localization \cite{ref2,ref3}. Furthermore, UAVs have been recently proposed to implement UAV-aided wireless communications, which have been identified as a solution for base-station offloading in extremely crowded areas, which is one of the 5G key scenarios \cite{ref4,ref5}. Important advantages of UAV-aided solutions are their flexibility and the absence of a fixed infrastructure, a particularly attractive characteristic for temporary, on-demand services and for disaster-recovery applications \cite{ref4}. Therefore, the implementation of measurements setups for Air-to-Ground (A2G) channel investigations is also important for the design of UAV-assisted communication and sensing systems. In addition, there are many emerging applications involving UAVs, therefore A2G channel characterization will provide valuable insights about physical properties of the radio channel for system designers.

\begin{figure*}[ht!]
\centering
\includegraphics[width=\textwidth]{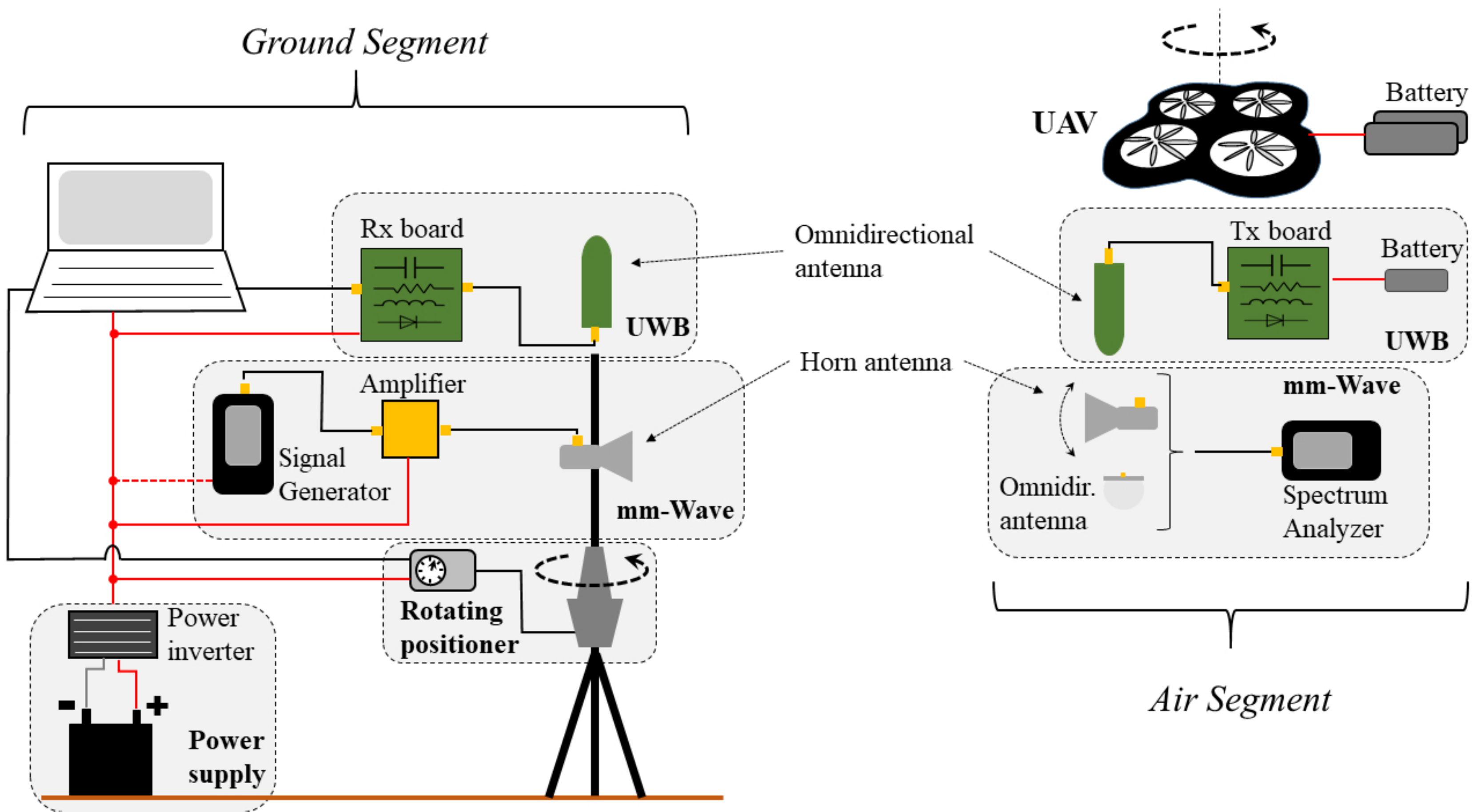}
\caption{Measurement set-up: air- and ground segment}
\label{fig:1}
\end{figure*}

Several experimental investigations have been carried out on the characterization of A2G propagation, especially over the last years \cite{ref6}. However, rural or open-field propagation \cite{ref7,ref8} and propagation in university campuses and sub-urban areas \cite{ref9,ref10,ref11,ref12} have been primarily addressed in many cases. Only a few studies have targeted A2G propagation in actual urban areas, probably due to the inherent difficulties in getting authorizations to fly on densely populated zones and/or close them to public during measurements \cite{ref13,ref14,ref15,ref16,ref17}. In some cases, the investigation has been limited to Air-to-Base-Station propagation \cite{ref9,ref16,ref17}, which is an interesting but less challenging scenario, compared to communications from the UAV to users at street level. Moreover, in such studies the analysis is mainly focused on large-scale parameters such as path-loss, fading statistics and spatial correlations, and is often limited to UHF or sub-millimeter wave frequencies. To the authors’ best knowledge, investigations on the millimeter wave channel, especially when aimed at its wideband or directional characteristics, have been usually carried out with terrestrial measurement setups \cite{ref18,ref19,ref20}, probably due to the problems involved in mounting and operating directive antennas or large MIMO transceivers on the UAV. Only a few investigations have considered the use of array antennas at either the ground station and/or the airborne station, and in this case, the array was mainly used to investigate performance of antenna diversity or low-order MIMO transmission schemes \cite{ref21,ref22,ref23}.

The present study aims at filling up some gaps in the field, and in particular at describing a flexible setup for double-directional wireless channel characterization, with focus on two millimeter-wave (mm-wave) frequencies, 27 and 38 GHz, that are quite popular for having been recently assigned to 5G systems. The proposed equipment and measurement campaign procedure is helpful to investigate low-altitude A2G propagation in urban areas, where the presence of buildings has a strong impact on the wireless communications. The measurement setup described in the present work is composed of a custom quadcopter equipped with a GPS and Real-Time Kinematic (RTK) localization/navigation system, and a mm-wave portable spectrum-analyzer connected to either a directive or an omnidirectional radiator. An Ultra-Wide Band (UWB) transceiver with UWB omnidirectional antenna that transmits carrier-less pulses and allows for the estimation of the channel’s impulse response is also mounted on the drone to measure the channel’s time-domain characteristics, albeit in the 3.1-5.3 GHz band. As the mm-wave setup is not phase-coherent, time-domain dispersion couldn’t be addressed. Nevertheless, since the multipath pattern spatial structure shouldn’t change significantly between different bands being the multipath pattern geometry the same, UWB results should give useful indications for the mm-wave channel as well. The ground station consists of the specular link-end, i.e. a mm-wave generator, an UWB transceiver and the corresponding antennas. Directive antennas are rotated in the azimuth plane at the ground station using a rotating positioner, while a servo-controlled gimbal is used on the drone for 3D steering capabilities.

The proposed measurement setup can be of interest for several applications, including the characterization of the directional properties of the A2G channel for beam-steering techniques, and the 3D, multi-band characterization of propagation in urban environment, with focus on mm-wave roof-to-street propagation, outdoor-to-indoor penetration, line-of-sight to non-line-of-sight transitions and the 3D characterization of scattering from buildings.

After the description of the customized measurement setup (Section \ref{sec:A2G_Meas_Setup}) and of the technical, logistical and legal issues related to the realization of the measurements campaign in a living urban area (Section ~\ref{sec:safety_auth_logistic}), some possible applications of the equipment are discussed in Section~\ref{sec:app_examples} together with some preliminary outcomes of UAV-assisted propagation measurements in urban environment. Conclusions are finally drawn in Section~\ref{sec:conclusion}.

\section{Air-to-Ground Measurement Set-Up}
\label{sec:A2G_Meas_Setup}
Air-to-ground wireless propagation has been investigated at both mm-wave and UWB frequencies by means of the experimental equipment outlined in Fig.~\ref{fig:1}. It consists of an air-segment and a ground-segment, as described in the following sub-sections. The air segment (drone) is a custom realization with unique features to reduce weight and to house in a limited space all the devices including a steerable mm-wave horn antenna with freedom of movement despite its rigid cable connection. Some insight into the UAV control features and the post-processing of the measured data is also provided.

\subsection{Air Segment}

\begin{figure}
\centering
\includegraphics[width=0.45\textwidth]{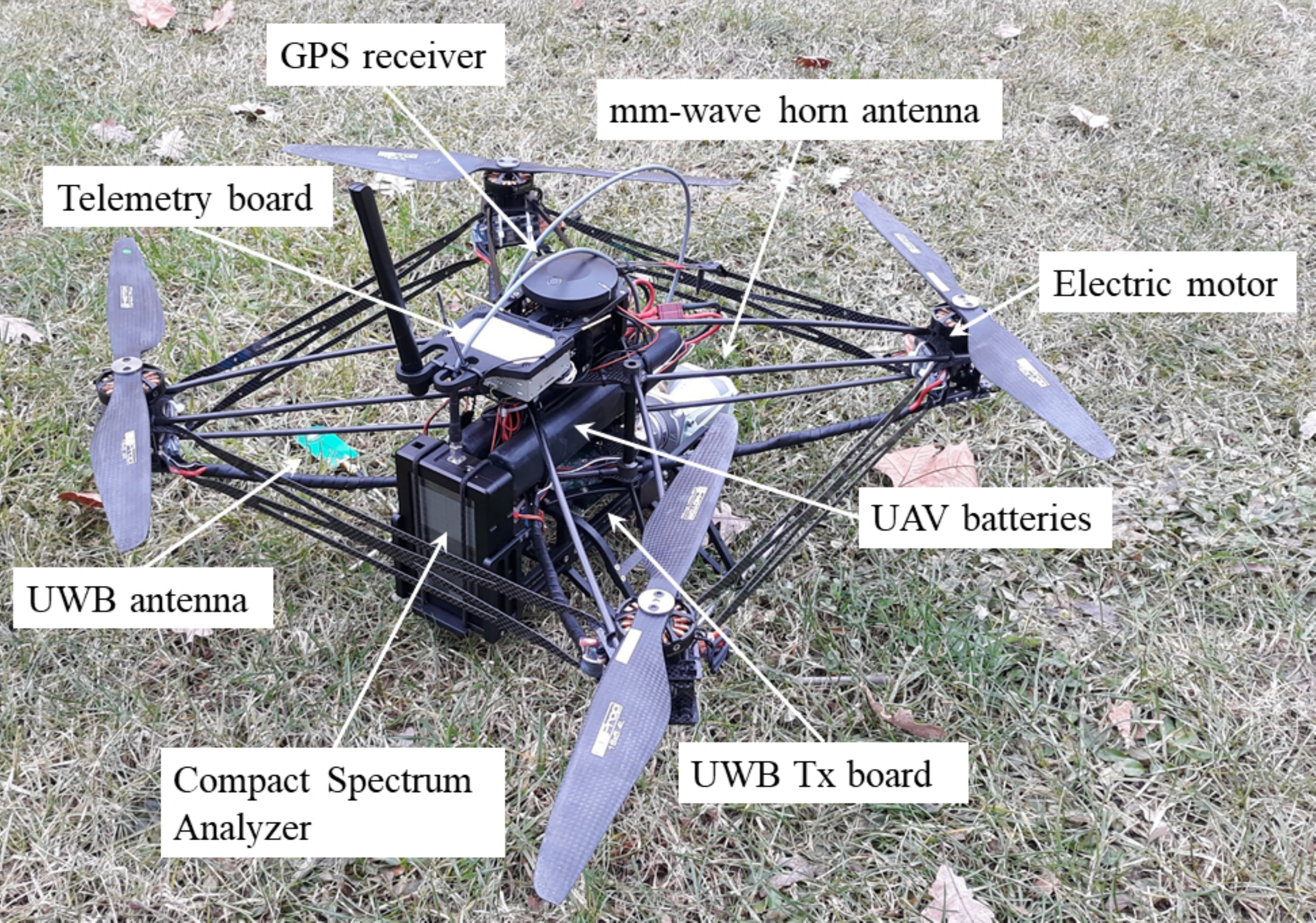}
\caption{Main components of the air station}
\label{fig:2}
\end{figure}

The UAV represents the backbone of the air station and consists of a quadcopter drone specifically designed by \cite{ref24} and customized for wireless channel measurement (Fig.~\ref{fig:2}). The drone frame is mainly made of carbon fiber, in order to limit the weight to 4~kg at most - measurement equipment included. A metal landing skid is present on the bottom side to let the UAV safely rest on ground, whereas the UAV control units \textendash{} e.g. the telemetry and remote-control boards and the GPS receiver \textendash{} are placed on the top.\newline

The major advantages of this custom design with respect to  a commercial drone are:
\begin{itemize}
\item	Use of open-source autopilot which allows to easy integrate the handling of the payload in a mission;

\item	Possibility to carry a 1.5~kg payload maintaining a 20~min flight time and with a max takeoff weight of 4~Kg to comply with latest drone regulation;

\item	Easier mechanical integration of the payload and weights balancing thanks to the availability and property of the whole CAD;
\end{itemize}
\bigskip

The main features of the custom UAV are briefly outlined as follows:\newline

\textbf{a)Frame}: the frame is the result of a custom design for applications of up to 1.5Kg payload. The main structure is composed of bent carbon fiber pipes which are kept in tension by carbon fiber plates at the tip. Other carbon fiber plates can be connected on the perimeter to increase the overall rigidity in case of high payload weight. The design was conceived to achieve:
\begin{itemize}
\item	Low weight: the overall frame weight (without electronics, powertrain and payload) is around 280g;

\item	High rigidity: the tensioned structure is very rigid and reduces the vibrations from the motors to the Inertial Measurement Unit;

\item	Aerodynamical efficiency: the use of thin carbon fiber pipes and plates reduces the aerodynamical drag of both the vertical flow of the propellers and the lateral wind disturbances;

\item	Quick and cheap reparation: hard crash usually cause the breaking of only a few carbon fiber pipes and plates on the arms which can be quickly and cheaply replaced;
\end{itemize} 
\bigskip

\textbf{b) Powertrain}: the powertrain was optimized to handle up to 1.5Kg of payload with a flight time of 20 minutes. It operates safety in a wide range of voltages (15V-25.2V) and at low currents to be compatible with Li-Ion 18650 cells which offer the best power density on the market. The motor used are T-motor MN4006 (380W each max power) and the propellers are T-motor 16x5.4 carbon fiber. The ESC is T-motor ALPHA esc 40.\newline

\begin{figure}
\centering
\includegraphics[width=0.45\textwidth]{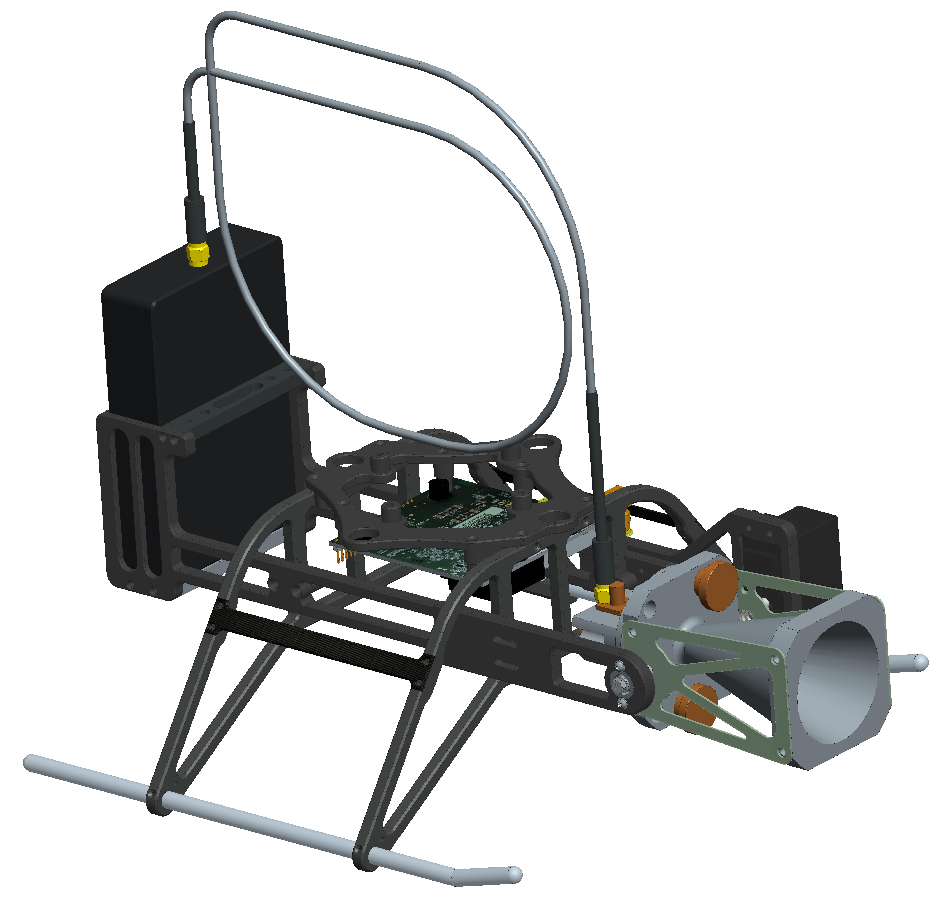}
\caption{CAD rendering of drone frame with payload and gimbal. Note the length and arrangement of the cable to allow for flexing.}
\label{fig:3}
\end{figure}

\textbf{c) Custom gimbal and payload}: the payload was aimed at effectively accommodating the antennas and the electronics. The horn antenna was shifted horizontally from the center of gravity to reduce interference with the drone structure. The weight shift was balanced by moving the spectrum analyzer in the opposite direction. Moreover, the antenna was mounted on a 1 axis gimbal to allow pitch orientation (to compensate the drone tilt and to perform different tests). The main problem in this case is to give the gimbal the needed freedom of movement with minimal resistance despite the presence of a very rigid mm-wave cable connection: a proper arrangement with a coil has been realized to solve this problem (see rendering of drone frame with gimbal and payload in Fig.~\ref{fig:3}).\newline

\textbf{d) Avionics}: the avionics include:
\begin{itemize}
\item	Autopilot board mRo Pixhawk 1;

\item	Hex Here+ RTK GPS module;

\item	Telemetry module: COFDM module @2.4GHz which offers a bidirectional transparent UART with maximum 921600 bps (used at 115200) as well as a FullHD 1080p video transmission to the ground station (not used yet). The safe maximum range of telemetry is around 2Km;

\item	Radiocontroller receiver: JETI duplex satellite 2.4EX to receive the commands from the pilot equipped with a JETI DS-12 transmitter;

\item	DCDC: commercial voltage converters to power the onboard electronics at 5V and 12V from the main battery of the drone;
\end{itemize} 
\bigskip

\textbf{e) Firmware}: among the different possible options, PX4 firmware v1.9.0 \cite{ref25} was finally selected, because of its features fairly well tailored to the requirements for payload and mission handling (e.g. simultaneous control of UAV rotation and antenna tilt through the servo control, see description below). \newline

The on-board measurement equipment consists of: i) mm-wave equipment and antennas; ii) UWB equipment and omnidirectional antenna, as described herein.\newline

\begin{figure}
\centering
\includegraphics[width=0.45\textwidth]{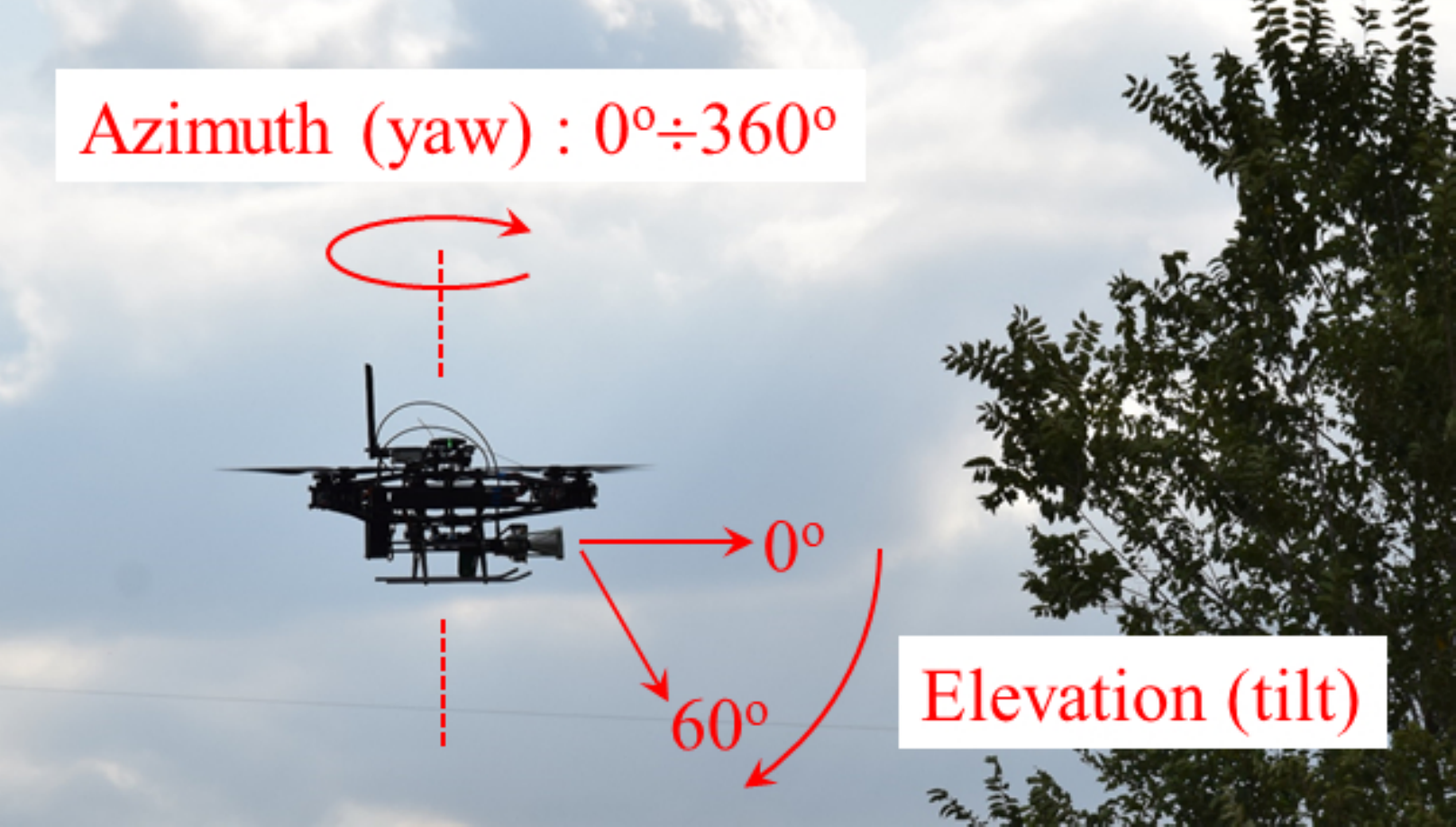}
\caption{Drone rotation angles for directional characterization of A2G propagation
}
\label{fig:4}
\end{figure}

\textbf{i) mm-wave equipment}: since drone authorization and registration procedures would have been much more complex if a RF transmitter in a licensed-band were onboard, the receiving mm-wave link-end has been placed on the drone, leaving the transmitting one to the ground station. A SAF Tehnika J0SSAP14 compact Spectrum Analyzer (SA) represents the air-end of the mm-wave measurement system \cite{ref26}. It can detect signal over the 26-40 GHz band with a sensitivity of about -100dBm. The sweep time is 0.5 sec. at the minimum frequency span, i.e. two Received Signal Strength (RSS) values per second can be recorded during the flight. Depending on the need, the SA can be fed by either a conical horn or an omnidirectional antenna, with gain respectively equal to about 21 dB and 3 dB over the operating frequency band. Polarization of the omnidirectional antenna is vertical, whereas it can be manually set to either vertical or horizontal for the horn antenna. Real-time control of the directive antenna elevation is achieved by means of the dedicated servo control, whereas its azimuthal pointing is managed by setting the yaw of the UAV during the flight. Because of mechanical constraints, the on-board antenna elevation can range from 0$^{\mathrm{o}}$ to 60$^{\mathrm{o}}$ downward (Fig.~\ref{fig:4}). In order to limit possible interference of the UAV frame on the antenna radiation properties, the horn antenna is offset from the drone center and always pointed outward, whereas the omnidirectional antenna is hanging from the UAV bottom side by a flexible cable tie. \newline

\textbf{ii) UWB equipment}: On-board equipment for pulse-radio UWB transmission consists of a light Humatics PulseON P410 radio transceiver board. During measurements, the P410 is configured for transmit mode. The transmitted signal spans on the 3.1-5.3 GHz frequency band and is radiated by an omnidirectional, vertically polarized antenna with a gain of about 3 dB (Humatics Broadspec UWB antenna) hanging from the UAV bottom side during flight, with proper distance from the drone’s body to avoid relevant scattering effects.

In order to save the drone battery life, the SA relies on its inner Li-ion batteries, as well as the UWB transmitter is supplied by a dedicated small Li-Po battery. The main features of the on-board devices are summarized in Table~\ref{tab:1}.

\subsection{Ground Segment}

\begin{figure}
\centering
\includegraphics[width=0.45\textwidth]{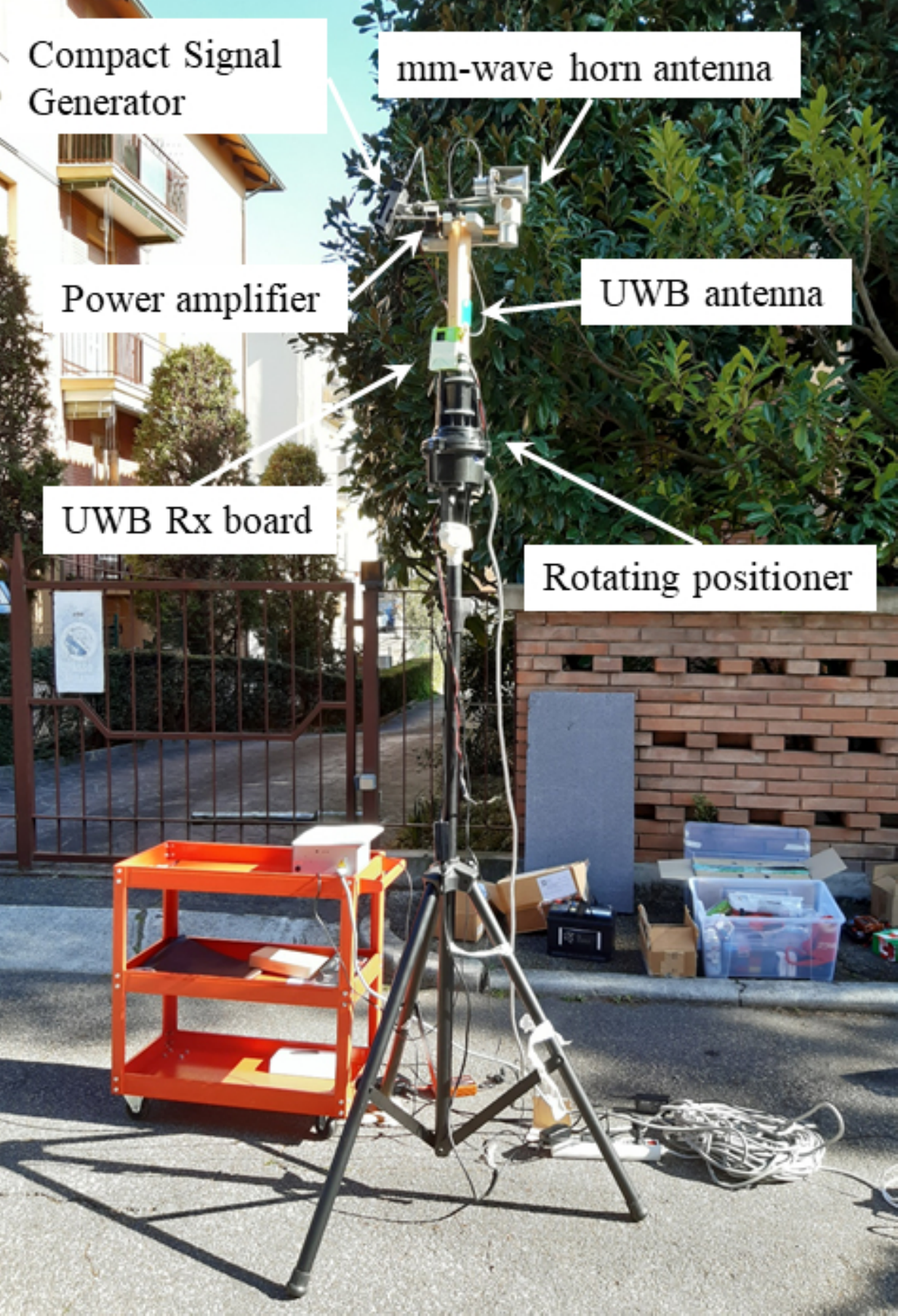}
\caption{Ground Station
}
\label{fig:5}
\end{figure}

At ground level, carrier wave signals in the mm-wave range are produced by a SAF Tehnika J0SSAG14 compact signal generator (SG) \cite{ref26}, boosted by a Ka-band power amplifier and eventually radiated by the same antennas already introduced in the previous section, now placed on the top of a portable tripod (Fig.~\ref{fig:5}). The SG maximum output power is equal to 5 dBm, and the amplifier has a gain of about 20 dB. In order to avoid overheating under hot weather conditions, the amplifier is provided with a fan-driven heat sink. Directional channel measurement can be carried out steering the horn antenna by means of a software controlled Yaesu g\hbox{-}450 rotating positioner (Fig.~\ref{fig:5}).

The UWB signal from the drone is collected at ground level by a second UWB omnidirectional antenna connected to a PC-controlled PulseON P410 board set in receiving mode (Fig.~\ref{fig:1}). Channel’s impulse responses are continuously estimated by the UWB control software and stored into the computer memory.

A power inverter connected to a 24V car battery provides the necessary power supply to the ground equipment, except for the SG that can be instead supplied by a Li-Po inner battery. The main information about the ground communication equipment is also listed in Table~\ref{tab:1}.

Finally, the ground station is completed by the ground end of the UAV control system, i.e. the telemetry ground unit and the radio control console operated by the pilot.

\subsection{Control System and Data Processing}

As highlighted in \cite{ref6}, wireless connections in the measurement setup consists of the payload communications and of the Control and Non-Payload Communications (CNPC) for the telematic management of the UAV. CNPC primarily includes telemetry and flight control, and employs the unlicensed 2.4 GHz band with a Coded Orthogonal Frequency-Division Multiplexing.

The full flight control is provided through the open source software QGroundControl (QGC) \cite{ref27}, which implements flight support for vehicles running PX4 or ArduPilot software (or any other autopilot based on the MAVLink protocol \cite{ref28}). QGC enables flight map display showing UAV position on the flight track, and real-time control of the on-board battery charge level, of the number of available satellites for GPS navigation and other flight data.

\begin{table*}
\caption{Main features of the Devices for UAV-to-ground channel measurements}
\label{tab:1}
\parnoteclear
\begin{tabularx}{\textwidth}{|
p{\dimexpr 0.09\linewidth-2\tabcolsep-2\arrayrulewidth}|
p{\dimexpr 0.16\linewidth-2\tabcolsep-\arrayrulewidth}|
p{\dimexpr 0.19\linewidth-2\tabcolsep-\arrayrulewidth}|
p{\dimexpr 0.12\linewidth-2\tabcolsep-\arrayrulewidth}|
p{\dimexpr 0.16\linewidth-2\tabcolsep-\arrayrulewidth}|
p{\dimexpr 0.28\linewidth-2\tabcolsep-\arrayrulewidth}|} \hline
 &  & \centering\arraybackslash{}\textbf{Make \& Model} & \centering\arraybackslash{}\textbf{Weight (g)} & \centering\arraybackslash{}\textbf{Power Supply} & \centering\arraybackslash{}\textbf{Main Features} \\\hline
\multirow{5}{=}{\textbf{Millimeter wave}}  & \textbf{Transmitter} \par (Ground segment) & SAF Tehnika J0SSAG14 \par Signal Generator & 400 & Li-Po 2200mAh inner battery, with life endurance ${\sim}$ 3h & Frequency range = 26-40 GHz \par Output power range = -3 ${\div}$ +5 dBm  \\\cline{2-6}
 & \textbf{Receiver} \par (Air segment) & SAF Tehnika J0SSAP14 \par Spectrum Analyzer & 400 & Two Li-ion 2380mAh inner batteries with life endurance ${\sim}$ 3h  & Frequency range = 26-40GHz \par Sensitivity = -100 dBm \par Resolution Bandwidth = 1 Mhz \par Sweep Speed = 0.5 (@ 100 MHz Span) \\\cline{2-6}
 & \textbf{Horn antenna} \par (Ground/Air segment) & SAF Tehnika  \par J0AA2640HG03 & 400 \textendash{} ground \par 330 \textendash{} on-board\parnote{In order to limit the payload, parts of the antenna supporting framework were removed when placed on the UAV} & \centering\arraybackslash{}--- & Operating frequency = 26.5-40.5 GHz \par Gain (typical) ${\sim}$ 21 dB \par Half Power Beam-Width: ${\sim}$ 12.5$^{\mathrm{o}}$ (E-plane), 15$^{\mathrm{o}}$ (H-plane)  \\\cline{2-6}
 & \textbf{Omnidir. Antenna} \par (Ground/Air segment) & Eravant \par SAO-2734030345-KF-S1 & 54 & \centering\arraybackslash{}--- & Operating frequency = 26.5-40 GHz \par Gain (typical) ${\sim}$ 3 dB \par Half Power Beam-Width (vertical) ${\sim}$ 45$^{\mathrm{o}}$ \\\cline{2-6}
 & \textbf{Amplifier} \par (Ground segment) & Eravant \par SBP-2734033020-KFK-S1 & 37 & 6-15V DC supplied by the power inverter in Fig.~\ref{fig:1} & Gain ${\sim}$ 20 dB \\\hline
\multirow{3}{=}{\textbf{UWB}}  & \textbf{Transmitter} \par (Air segment) & \multirow{2}{=}{\parbox{\textwidth}{Humatics\\ PulseON P410}}  & \multirow{2}{=}{58}  & Li-Po battery - on-board & \multirow{2}{=}{Signal bandwidth = 3.1-5.3 GHz}  \\\cline{2-2}\cline{5-5}
 & \textbf{Receiver} \par (Ground segment) &  &  & 5.75-30V DC supplied by the power inverter in Fig. 1 \textendash{} at ground &  \\\cline{2-6}
 & \textbf{Omnidir. Antenna} \par (Ground/Air segment) & Humatics BroadSpec$^{\mathrm{TM}}$ UWB Antenna & 10 & \centering\arraybackslash{}--- & Gain ${\sim}$ 3 dB \\\hline
\end{tabularx}
\parnotes
\end{table*}

Moreover, QGC also supports the planning of flight missions, encoded as a list of interleaved ``waypoints'' (WP) and ``regions of interest'' (ROI). WPs set the spatial positions the UAV will fly through, and consist of the corresponding coordinates (latitude, longitude and altitude above ground level) and a holding time the UAV will spend hovering on the WP. Conversely, ROIs identify the points the on-board antenna must point at along the mission. The UAV yaw and the onboard servo control are therefore arranged according to both the current ROI and the UAV position. Planned missions can be uploaded to the UAV for execution through the CNPC link. A C\# Graphical User Interface has also been created, in order to automate and facilitate mission planning.

QGC also manages telemetry record logs (Tlogs), which are recording of the MAVLink telemetry messages including a wide range of information about the UAV, like its position and yaw, the steering angle of the servo control, etc. Tlogs are created when the UAV takes-off and are stored at landing. They are finally downloaded through the CNPC or via a removable SD card.

Synchronization between the QGC software and the measurement equipment is needed to effectively match payload data (propagation measurements data) with CNPC (telemetry information) afterwards. In practice, this is accomplished by setting the inner clocks to the Universal Time Coordinated (UTC). It is worth noticing that, due to UAV holding time in the same position always equals a few seconds, a very precise synchronization is not necessary (synchronization of the order of the second is sufficient).

In spite of the autopilot, the presence of an expert pilot is fundamental as far as missions are carried out in critical areas like dense urban environments, where GPS/RTK might be not always reliable, especially during take-off and landing, because satellites may be hidden by buildings or GPS signals somehow corrupted by multipath effects.

Finally, \textit{ad hoc} software has been developed for measurement data analysis. Different data processing procedures must be carried out depending on the mission type (see Section IV). When the drone hovers in a fixed position with the same antenna pointing (e.g. for roof-to-street measurement, Section~\ref{sec:app_examples}), CNCP data are unnecessary as they don’t account for the movement of the ground station, which was instead manually recorded during the mission. Conversely, telemetry information are fundamental to get A2G channel characterization whenever the drone travels on flying paths and/or the on-board antenna is steered in different directions. In such a case, measured data are related to the UAV position/orientation by matching the corresponding time stamps. As the UAV hovering time on the different WPs is of 5 s, at least 10 times (mm-wave setup) and 50 times (UWB setup) greater than the sweep time of the wireless receivers, time averaging of the RSS is always performed to limit possible fluctuations due to the imperfect stability of the drone. In alternative, sampling and analysis of fast fading could be performed using a different procedure, for example by performing continuous flights over small circles around measurement locations and by collecting data without any averaging.\newline

\section{Safety, Authorization and Logistics}
\label{sec:safety_auth_logistic}

This section describes some of the authorization and logistic issues that must be addressed to accomplish drone-based urban measurements. Although the limitations and the requirements for UAV operations may currently differ at national level, the regulation enforced by the Italian civil aviation Authority (ENAC) \cite{ref29} \textendash{} which is the main reference herein - is almost compliant with the European harmonized regulation that is fully applicable from December 30th, 2020 \cite{ref30}.

The first step toward the realization of the measurement campaign has been the choice of the locations. Although a civil drone fly permit can be issued for almost all areas upon application and specific evaluation by ENAC, the authorization process turns out faster if flying is limited to 50~m above-ground and outside of approach and landing corridors of civil and military airports. The mid-sized town of Imola (IT) was therefore selected, as it is far enough from surrounding airports and, being relatively small, has uncrowded areas even in the urban core that can be closed to traffic without much inconvenience for local residents and with a relatively easy logistic and authorization procedure.

The design of the drone and of the payload has represented the second step of the experimental activity. A total weight not exceeding 4 kg was set as a crucial requirement, as it is the limit for ``class A2'' drones, that must comply with relatively simpler safety restrictions and rules compared to heavier drones and fixed-wing UAVs of higher class. The main requirements to perform class A2 drones missions are shortly outlined in the following list:

\begin{itemize}
\item a flight permit issued by ENAC upon detailed mission description and declaration of ``non-critical'' drone operations;

\item closure of the area underneath the flight zone to traffic and people with a keep-off margin of 30 m at ground level below the UAV;

\item a pilot with an official license;

\item mandatory registration of the drone, of the pilot and of the ``drone operator'', i.e. the person or company promoting, and responsible for, the mission;

\item mandatory on-board identification through a QR code sticker;

\item mandatory insurance;

\item mandatory risk analysis for each single drone mission to be filed beforehand;

\end{itemize}

Further application for authorization must be also submitted to the local administration some days before any drone mission, including a detailed description of the drone flights and proof of compliance with the previous requirements.

\begin{figure}
\centering
\includegraphics[width=0.45\textwidth]{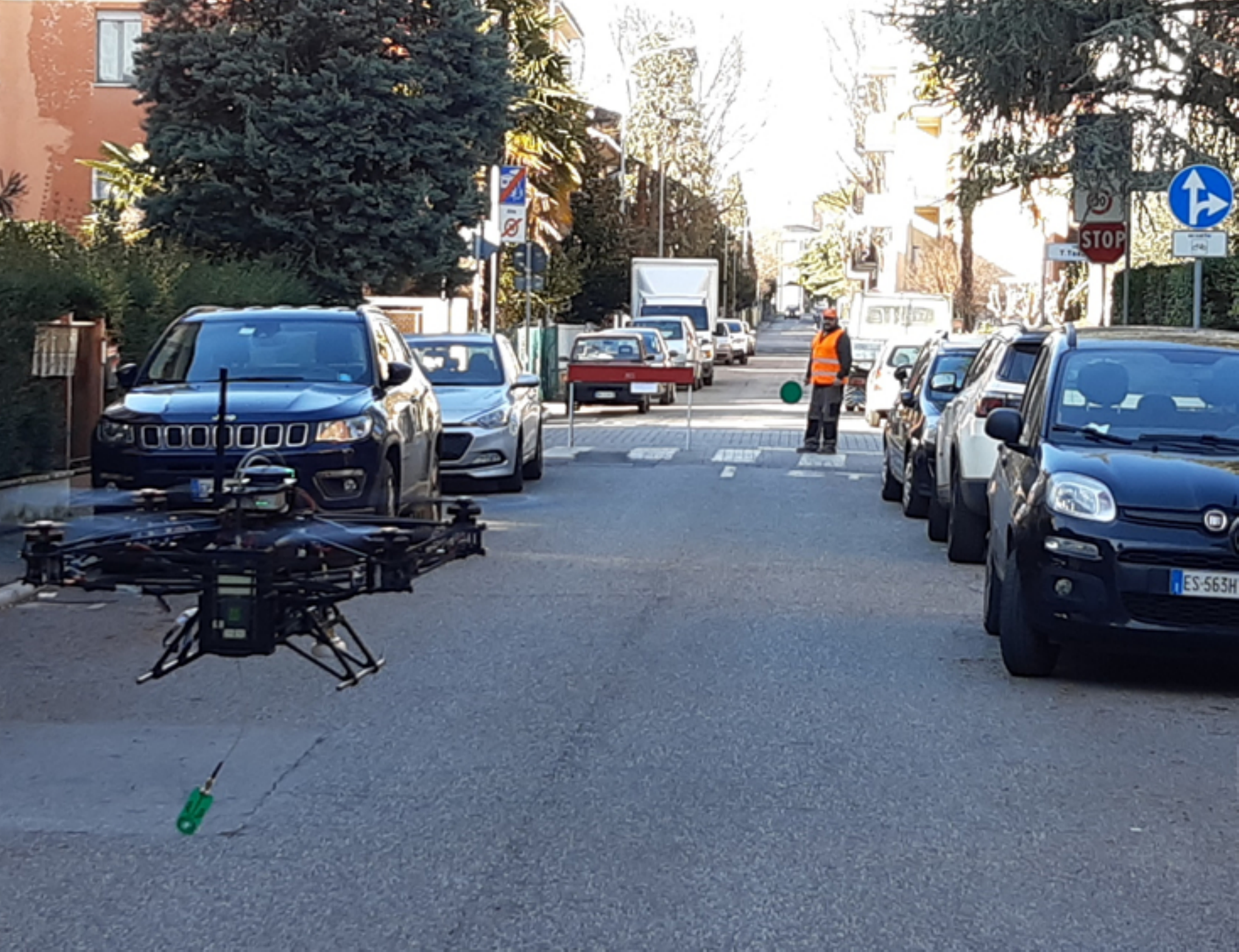}
\caption{Measurement activity in urban environment
}
\label{fig:6}
\end{figure}

During the measurement activities, temporary traffic restrictions were enforced with the help of professional signaling personnel using no-entry signs and barriers to dissuade vehicles and pedestrians from entering the keep-off area. Inhabitants of buildings inside the area were warned to stay indoor during flights (Fig.~\ref{fig:6}).

Flight routes had to be designed to ensure full visibility to the pilot, who can promptly disable the autopilot and take manual control of the UAV in case it gets dangerously close to some obstacles or starts flying awkwardly, because of GPS unreliability in urban environment or other technical problems. In case of extreme emergency, the pilot is empowered to give the "kill command" through a dedicated switch on the drone control, which simply switch the engines off thus crashing the UAV to the ground.

Of course, rainy, low-visibility or windy days had to be discarded from possible measurement days.

\section{Application Examples}
\label{sec:app_examples}
\begin{figure}
\centering
\includegraphics[width=0.5\textwidth]{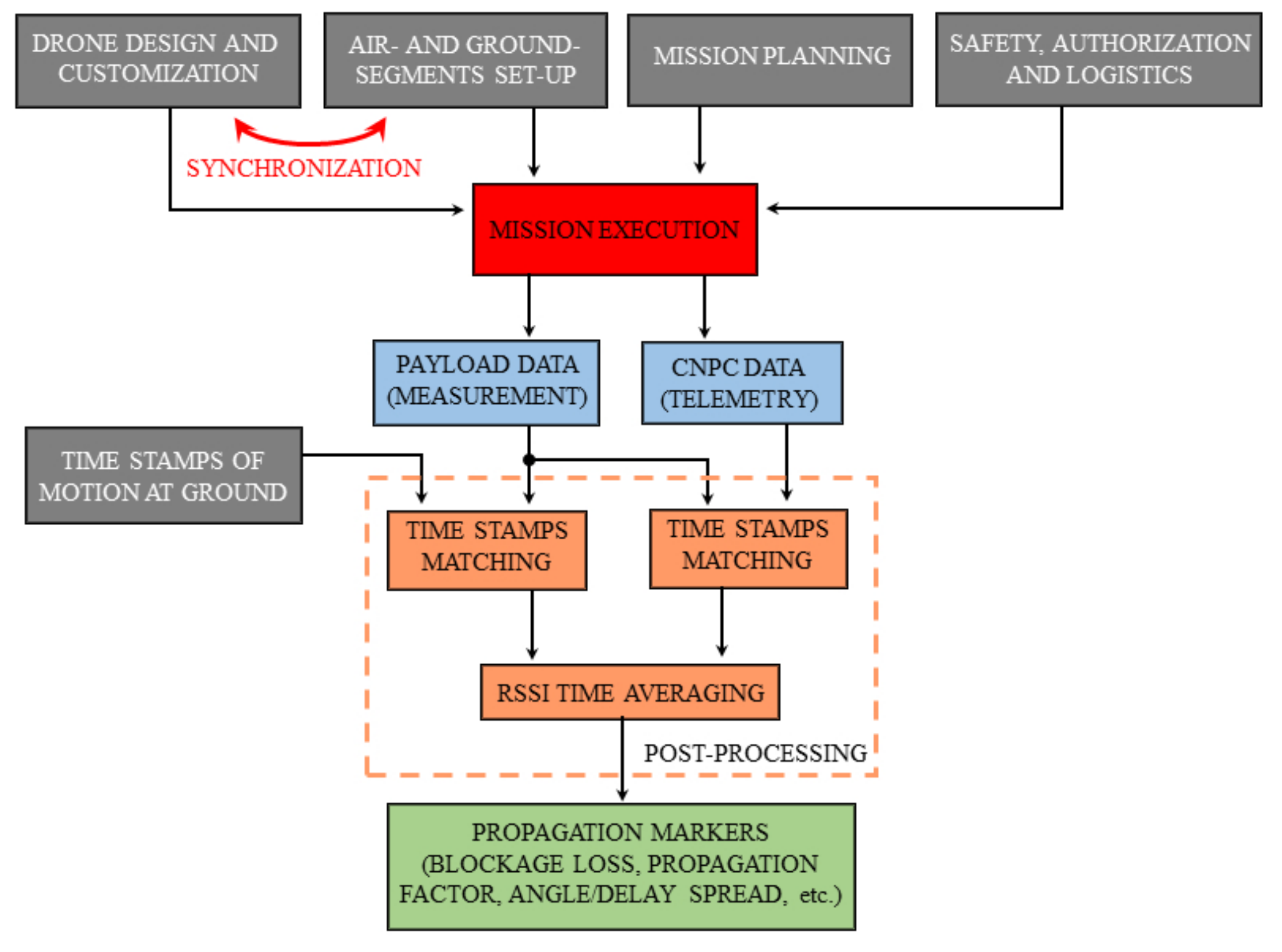}
\caption{Flow chart of the overall measurement process
}
\label{fig:7}
\end{figure}

Thanks to its flexibility, the measurement setup described in the previous sections can be useful in several practical cases, briefly outlined herein. Regardless of the specific application, measurement planning, execution and data post-processing has always been carried out according to the general procedure outlined by the flow chart in Fig.~\ref{fig:7}. As already referred to at the end of Section ~\ref{sec:A2G_Meas_Setup}C and also highlighted by the chart, time synchronization of the different components of the measurement system (mm-wave and UWB equipment, both onboard and at ground) with the UAV telemetry system may represent a key issue and therefore, synchronization with GPS time is carried out before starting every mission. This allows precise matching of the measurement and telemetry timestamps during the post-processing stage, so that every collected sample (RSS value, Power-Delay Profile) can be associated with the current WP and ROI recorded into the telemetry logs. This means that, eventually, each measured sample can be associated with the UAV GPS coordinates (retrieved from the WP), while the azimuthal pointing and tilt angles of the horn antenna can be retrieved from the recorded ROI orientation with respect to the current WP. Moreover, time averaging of the data samples corresponding to the same WP and ROI has been performed, in order to compensate for fading and unavoidable UAV vibrations, as already pointed out. As takeoff and landing are operated manually by the pilot, all measurement data collected before and after the automated mission are also discarded during the post-processing stage.

The following subsections provide some examples of different measurement tasks and applications, in order to show the full potential and flexibility of the proposed setup.

\subsection{Full-3D measurements in the angular domain}

By exploiting the directive horn antenna and the drone rotation capability in the azimuth domain (``yaw'' angle in the drone telemetry), Power-Azimuth Profiles (PAP) can be obtained with straightforward post-processing, i.e. synchronization of telemetry and measurement data. Moreover, as the horn antenna is driven by the servo control, different elevation (tilt) angles of the on board antenna can be considered, in order to get a full-3D Power-Angle-Profile at the air station (Fig.~\ref{fig:4}). As the ground station is also equipped with a rotating positioner, in the end a double-directional characterization of the A2G channel can be achieved.

\begin{figure}
\centering
\includegraphics[width=0.45\textwidth]{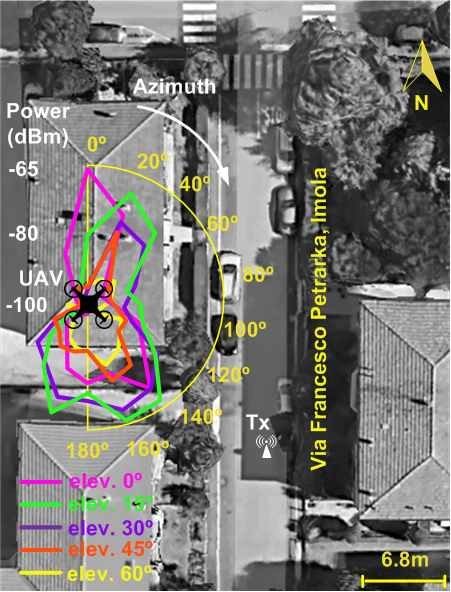}
\caption{Top-view of the street-canyon scenario and corresponding PAPs for different tilt angles.}
\label{fig:8}
\end{figure}

An example of 3D Power-Angle Profile at the air station is reported in Fig.~\ref{fig:8}, corresponding to a street-canyon scenario with the air station located few meters above the building roof (19 m from ground level), while the ground station is placed at 2 m height at the opposite side of the street, close to the sidewalk. The operating frequency is 27 GHz. This is a quasi-Line of Sight (LOS) scenario, as the roof partially shadows the main lobe of the on board antenna, when it is pointing towards the ground station. In Fig.~\ref{fig:8}, curves in different colors represent the PAP for the different elevation of the UAV antenna, ranging from 0$^{\circ}$ to 60$^{\circ}$ (downwards) with 15$^{\mathrm{o}}$ step. Each azimuth sample corresponds to a different drone yaw with respect to North, from 0$^{\circ}$ to 360$^{\circ}$ (clockwise), still with 15$^{\circ}$ step. Looking at the figure, the optimum tilt angle seems to be equal to 15$^{\circ}$, whereas the RSS decreases for greater tilt angles, as the horn antenna is increasingly pointing toward the roof. Moreover, some ``secondary lobes'' at 0$^{\circ}$${\div}$20$^{\circ}$ and 200$^{\circ}$${\div}$220$^{\circ}$ are clearly visible in Fig.~\ref{fig:8}, in addition to the LOS contribution at 140$^{\circ}$: these are probably caused by reflection and scattering from the vertical walls of taller buildings located around.

\begin{figure}
\centering
\includegraphics[width=0.45\textwidth]{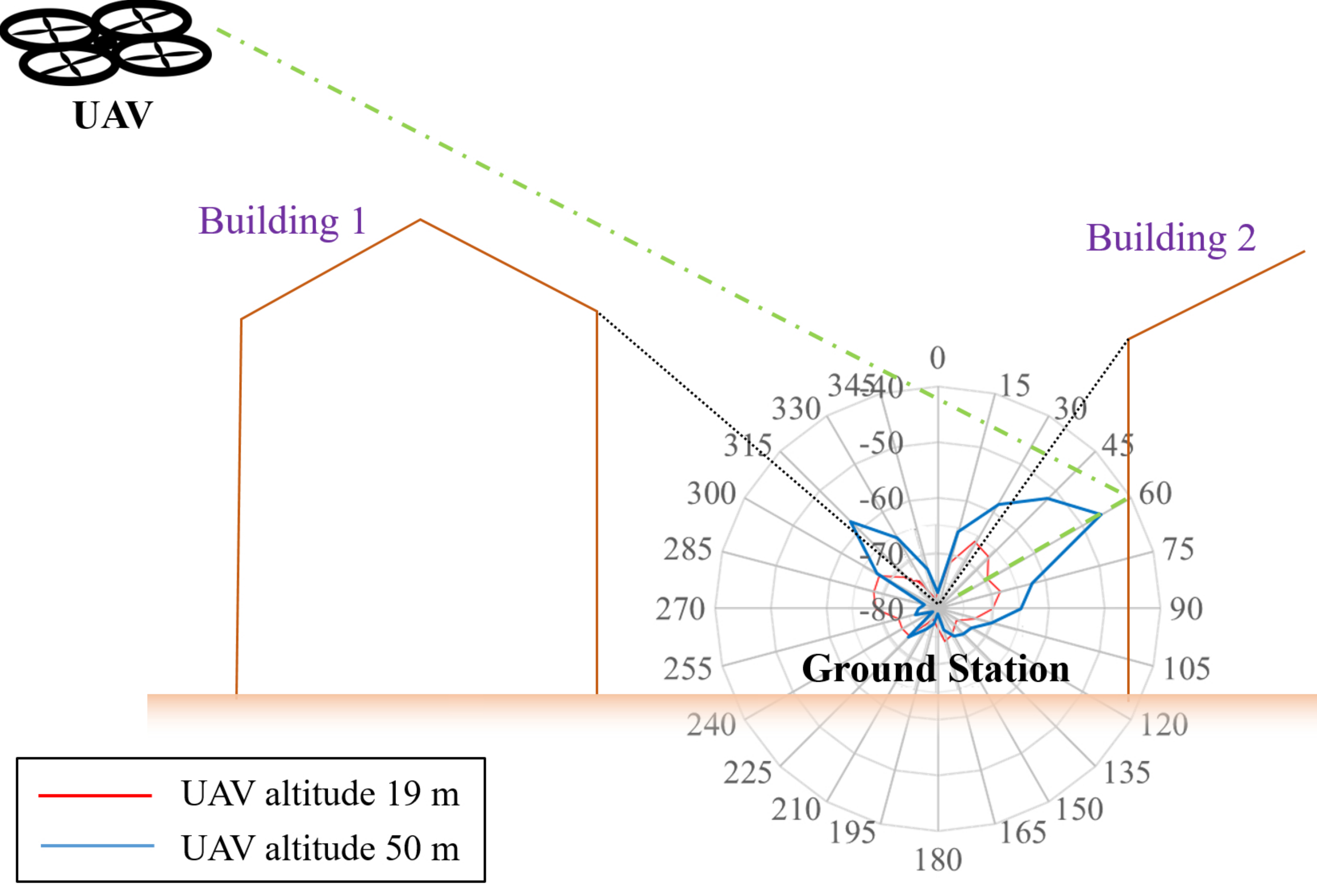}
\caption{Power-Elevation Profiles at TX in the "Air-to-street" scenario, for 2 different drone altitudes: 19 m (red curve) and 50 m (blue curve).}
\label{fig:9}
\end{figure}

A further example of characterization in the angular domain is the ``air-to-street'' propagation scenario, where the ground station is placed on one side of an urban street canyon, while the UAV is hovering in a fixed position above the roof of a building on the opposite side of the street. In such a case, the directive antenna is mounted on the rotating positioner at the ground station, and the omnidirectional antenna is mounted on the UAV, and the frequency is the same as above (27 GHz). The measured Power-Elevation Profile (PEP) at the ground station is shown in Fig.~\ref{fig:9} for two different altitudes of the air station (19 m and 50 m). As shown in the figure, since the direct path between UAV and Ground Station is obstructed by Building 1, there are 2 possible propagation mechanisms allowing the transmitted signal from the UAV to reach the RX in the street canyon: diffraction on the roof of Building 1, and reflection/scattering from the building behind Rx (Building 2). This result shows that for the higher altitude (blue line in Fig.~\ref{fig:9}), the dominant propagation mechanisms is the specular reflection on Building 2, in agreement with previous studies \cite{ref31,ref32}.

The results shown in this subsection have been partly discussed also in \cite{ref33}.

\subsection{Horizontal and vertical scans at different altitudes}

Thanks to the drone mobility, horizontal and vertical scans can be easily obtained in a number of different cases. For example, power or path loss can be computed along a street canyon at different heights, or LOS/Non LOS transitions can be analyzed when the UAV is crossing the main street, or when it is flying beyond the rooftop of a building facing the street. The use of the omnidirectional antenna at the UAV side is more suitable for this kind of measurement. Besides, information in the delay domain (e.g. RMS Delay Spread) or about the Doppler shifts can be also retrieved by means of the UWB equipment.

\begin{figure}
\centering
\includegraphics[width=0.45\textwidth]{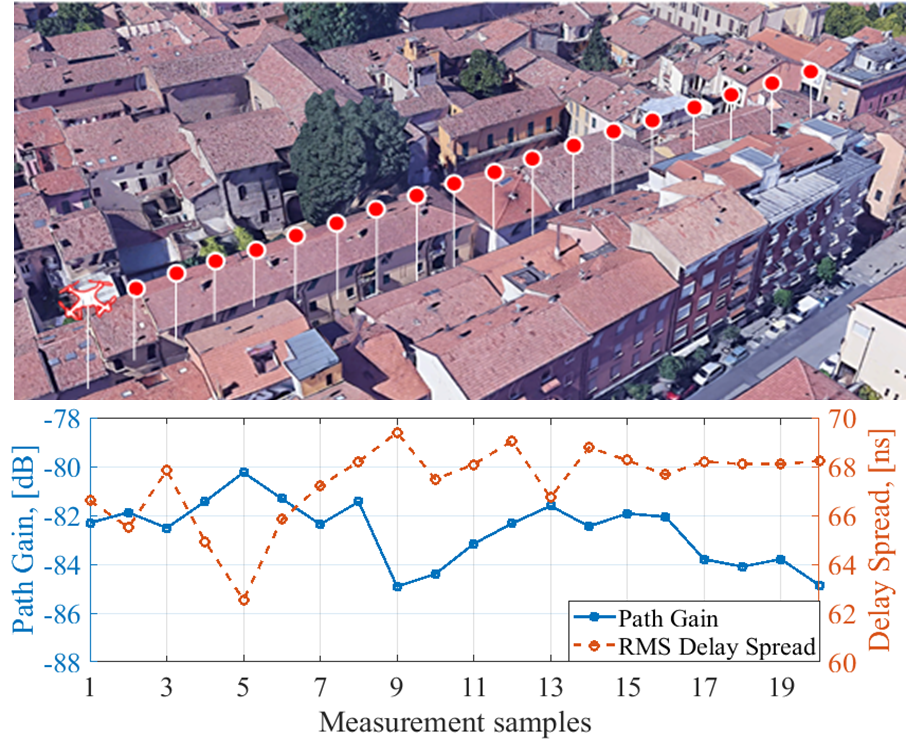}
\caption{UWB measurement along an urban street canyon at 16 m altitude: Path Gain and RMS Delay Spread are reported for each measurement location}
\label{fig:10}
\end{figure}

In Fig.~\ref{fig:10}, an example of Path Gain and Delay Spread measured in the UWB band along an urban street canyon is reported, for 20 static measurement points at a distance of 5m with the air station located in the middle of the street at a height slightly higher than the average building height (16 m), while the ground station is placed at the beginning of the street (on the left in the picture) at a distance of about 100 m from the start of the measurement route. It can be noted that while Path Gain slowly decreases with distance, as expected, Delay Spread instead increases.

\subsection{3D Scattering from buildings}

\begin{figure}
\centering
\includegraphics[width=0.45\textwidth]{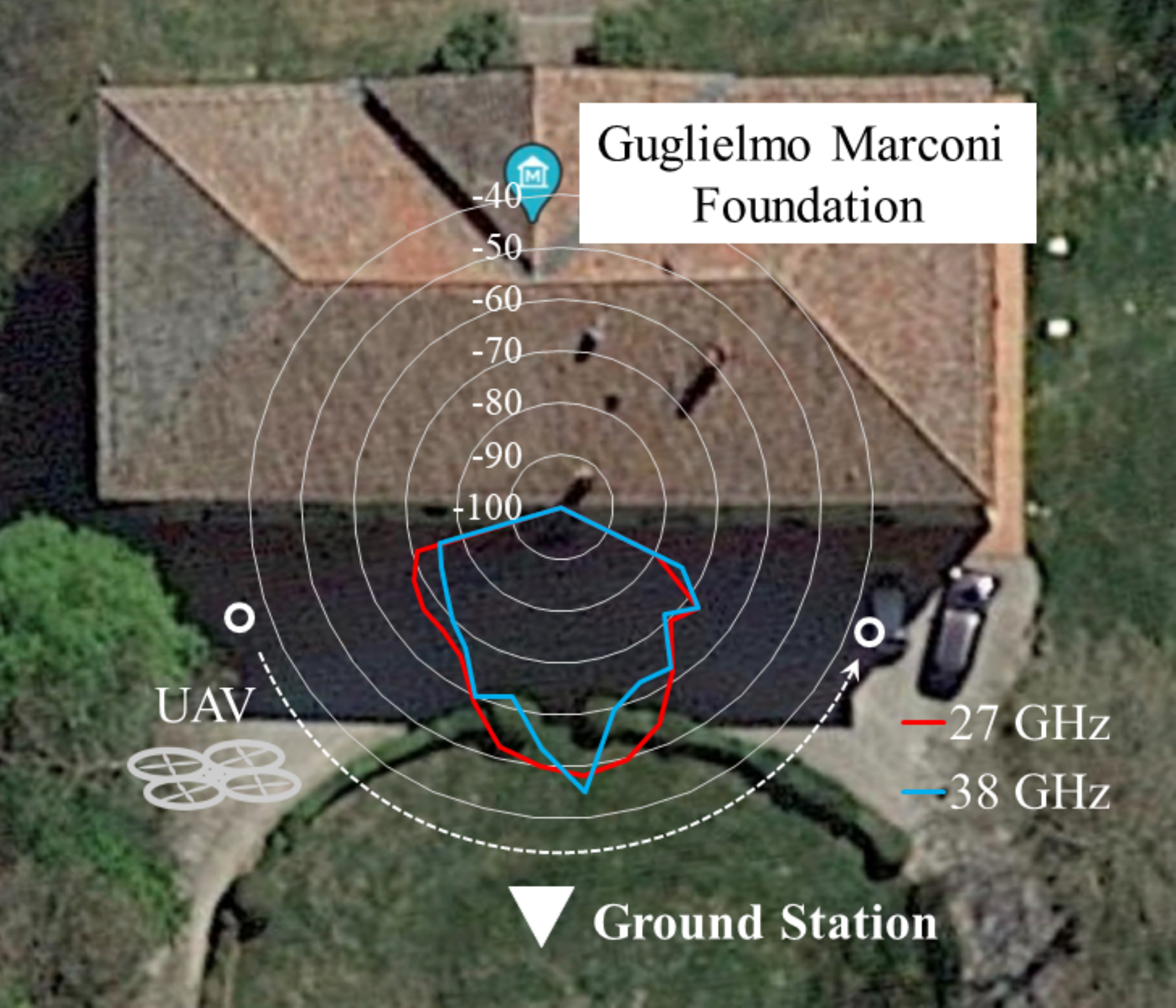}
\caption{Horizontal pattern of the back-scattering from a building wall, corresponding to the UAV flying in the azimuth plane}
\label{fig:11}
\end{figure}

Using directive antennas at both link ends (drone and ground station), 3D directional information about the power backscattered from a building wall can be achieved. Such measurement can be performed with the ground antenna illuminating a fixed ``spot'' on the building fa\c{c}ade, while the UAV moves over a semicircle in front of the building keeping the on board antenna turned in the direction of the spot (Fig.~\ref{fig:11}).

\begin{figure}
\centering
\includegraphics[width=0.45\textwidth]{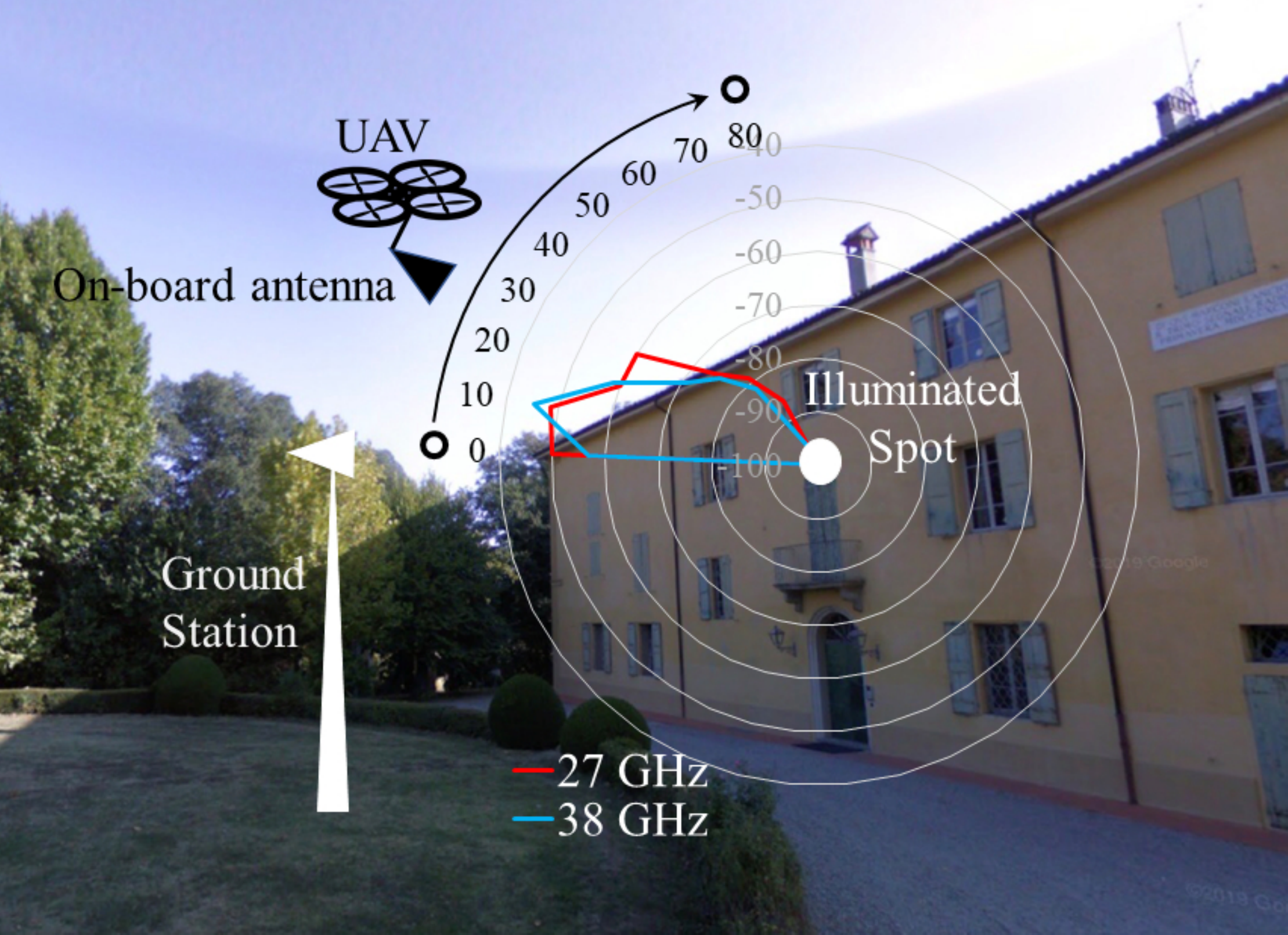}
\caption{Vertical pattern of the back-scattering from a building wall, corresponding to the UAV flying in the elevation plane}
\label{fig:12}
\end{figure}

Differently from previous works where backscattering was analyzed in the horizontal plane only \cite{ref34}, the UAV-based setup allows to obtain a scattering pattern also in the vertical plane, where the on-board antenna is required to reach a height of the order of tens of meters, that would be indeed very difficult with a traditional ground-based measurement system. Fig.~\ref{fig:11} and Fig.~\ref{fig:12} show an example of horizontal and vertical scattering patterns of an historical building (Villa Griffone in the outskirt of Bologna - IT - where Guglielmo Marconi carried out his first experiments), with the ground antenna placed at 10 m in front of the building and the UAV flying over a circular path at 8 m from the target spot.

\subsection{Above Ground Level and Outdoor-to-Indoor coverage measurements}

Another great advantage of the measurement setup described in this work consists of the possibility of making the UAV fly along the fa\c{c}ades of buildings, in order to investigate outdoor coverage at high altitudes, i.e. Above Ground Level (AGL), from a Base Station (BS) located nearby. By doing so, many limitations of traditional measurement setups can be overcome, as outdoor AGL measurement are usually possible to a very limited extent, e.g. by using a probe on windows or balconies, if available \cite{ref35}.

These outdoor measurements can be then complemented with further coverage measurements performed indoor at different floors of the building, so that the effects of outdoor-to-outdoor and outdoor-to-indoor (O2I) propagation are analyzed separately. Indoor measurements can be carried out either using a traditional ground-based measurement system which is moved to different locations inside the building through a trolley, or using a second UAV, for example a DJI Phantom Drone of smaller size and lower weight, therefore more suitable for indoor measurements \cite{ref36}. Finally, the (average) building penetration loss can be computed as the difference between the RSS values collected outdoor and indoor.

\subsection{Emerging applications and future prospects}

As discussed above, UAV-based measurement setups can be formidable tools to perform full-3D propagation characterization. However, ground-based, fully coherent MIMO channel sounders can resolve even minor multipath components in both time and angles, which is a crucial feature, especially at mm-wave frequencies. Building similar channel sounders for UAV-based setups is a challenge due to multiple factors. At mm-wave frequencies, due to the short wavelength, any drift of the drone can cause a severe phase change. Unless some tethered solution is considered, which still limits the use of the measurement system in urban environment, it is very difficult to synchronize UAV and ground station in order to get accurate phase information. Moreover, any vibrations present at the UAV can cause severe phase noise and discrepancies. Our ongoing work includes trying to find solutions to the mentioned problems and developing more accurate and complete UAV-based setups for future measurement campaigns.

\section{Conclusion}
\label{sec:conclusion}
A drone-based measurement setup for the characterization of wireless propagation in urban environment has been presented in this work. Compared to traditional ground-based measurement systems, the proposed equipment offers much greater flexibility and versatility, as the drone can be easily placed and moved almost everywhere within the urban context, from ground level to the top of buildings and beyond.

At the same time, flying a drone in urban environment represents a rather challenging task, with manifold issues to take care of. First, the drone must be specifically design to carry the measurement equipment, which has to be as light and compact as possible and must be effectively placed in order to limit possible interference rising from the drone frame. Furthermore, the measurement site as well as the flying routes have to be carefully selected beforehand according to strict safety requirements. In this respect, both the drone design and the flight plans need to be approved by entitled authorities and institutions. Finally, measurements execution requires the closure of the area underneath the drone to traffic and pedestrians, and the presence of a qualified pilot to take manual control of the drone in case of possible unreliability of the autopilot, e.g. for GPS inaccuracies in urban environment.

The proposed measurement setup can be employed in many different investigations, including the directional characterization of the air-to-ground channel, the assessment of above ground level and outdoor-to-indoor propagation, and the evaluation of line-of-sight to non-line-of-sight transitions and of the 3D scattering pattern of buildings, as shown in section IV. The use of this measurement setup and procedure for a thorough characterization of air-to-ground propagation in urban environment will be object of future work.

\section*{Acknowledgment}
Authors would like to thank Dr. Michele Furci and Mr. Andrea Sala, from the AslaTech company, for the UAV design and customization, as well as for the support to address authorization to fly and flight safety issues throughout the measurement activity.

\bibliographystyle{IEEEtran}

\begin{thebibliography}{9}

\bibitem{ref1} D. W. Matolak and R. Sun, ``Unmanned Aircraft Systems: Air-Ground Channel Characterization for Future Applications'', \textit{IEEE Vehic. Tech. Magazine}, vol.10, no. 2, June 2015, pp. 79\textendash{}85, DOI: \href{https://doi.org/10.1109/MVT.2015.2411191}{10.1109/MVT.2015.2411191}.

\bibitem{ref2} C. Li, E. Tanghe, D. Plets, P. Suanet, J. Hoebeke, E.D. Poorter, W. Joseph, "ReLoc: Hybrid RSSI- and Phase-Based Relative UHF-RFID Tag Localization With COTS Devices", \textit{IEEE Trans. on Instr. and Meas.}, vol. 69, no. 10, pp. 8613-8627, October 2020, DOI: \href{https://doi.org/10.1109/TIM.2020.2991564}{10.1109/TIM.2020.2991564}.

\bibitem{ref3} E.A. Zhidko, S.N. Razin’kov, ``Methods for Determining the Angular Coordinates and Locations of Radio Sources in Unmanned Monitoring Systems and Experimental Estimates of the Accuracy of these Parameters'', \textit{Measurement Techniques}, vol. 62, February 2020, pp. 893-899, DOI: \href{https://doi.org/10.1007/s11018-020-01710-6}{10.1007/s11018-020-01710-6}.

\bibitem{ref4} Y. Zeng, R. Zhang and T. J. Lim, "Wireless communications with unmanned aerial vehicles: opportunities and challenges," \textit{IEEE Comm. Magazine}, vol. 54, no. 5, pp. 36-42, May 2016, DOI: \href{https://doi.org/10.1109/MCOM.2016.7470933}{10.1109/MCOM.2016.7470933}.

\bibitem{ref5} D. He, S. Chan and M. Guizani, "Drone-Assisted Public Safety Networks: The Security Aspect," \textit{IEEE Comm. Magazine}, vol. 55, no. 8, pp. 218-223, Aug. 2017, DOI: \href{https://doi.org/10.1109/MCOM.2017.1600799CM}{10.1109/MCOM.2017.1600799CM}.

\bibitem{ref6} W. Khawaja, I. Guvenc, D. W. Matolak, U. Fiebig and N. Schneckenburger, "A Survey of Air-to-Ground Propagation Channel Modeling for Unmanned Aerial Vehicles," \textit{IEEE Comm. Surveys \& Tutorials}, vol. 21, no. 3, pp. 2361-2391, third quarter 2019, DOI: \href{https://doi.org/10.1109/COMST.2019.2915069}{10.1109/COMST.2019.2915069}.

\bibitem{ref7} Z. Cui, C. Briso-Rodriguez, K. Guan and Z. Zhong, "Ultra-Wideband Air-to-Ground Channel Measurements and Modeling in Hilly Environment," \textit{2020 IEEE Int. Conf. on Communications} (ICC), Dublin, Ireland, June 7-11, 2020.

\bibitem{ref8} W. Khawaja, O. Ozdemir, F. Erden, I. Guvenc and D. W. Matolak, "Ultra-Wideband Air-to-Ground Propagation Channel Characterization in an Open Area," \textit{IEEE Trans. on Aerospace and Electronic Systems} (early access), June 2020, DOI: \href{https://doi.org/10.1109/TAES.2020.3003104}{10.1109/TAES.2020.3003104}.

\bibitem{ref9} X. Cai et al., "An Empirical Air-to-Ground Channel Model Based on Passive Measurements in LTE," \textit{IEEE Trans. on Veh. Tech.}, vol. 68, no. 2, pp. 1140-1154, Feb. 2019, DOI: . \href{https://doi.org/10.1109/TVT.2018.2886961}{10.1109/TVT.2018.2886961}.

\bibitem{ref10} Zhihong Qiu, Xi Chu, Cesar Calvo-Ramirez, Cesar Briso, and Xuefeng Yin "Low Altitude UAV Air-to-Ground Channel Measurement and Modeling in Semiurban Environments", \textit{Wireless Comm. and Mobile Computing}, Hindawi, 2017, DOI: \href{https://doi.org/10.1155/2017/1587412}{10.1155/2017/1587412}.

\bibitem{ref11} Z. Cui, C. Briso-Rodriguez, K. Guan, Z. Zhong and F. Quitin, "Multi-Frequency Air-to-Ground Channel Measurements and Analysis for UAV Communication Systems," \textit{IEEE Access}, vol. 8, pp. 110565-110574, 2020, DOI: \href{https://doi.org/10.1109/ACCESS.2020.2999659}{10.1109/ACCESS.2020.2999659}.

\bibitem{ref12} M. Gauger, M. Arnold and S. ten Brink, "Drone-Based Spatial MIMO Measurements in Three Dimensions," \textit{24th Int. ITG Workshop on Smart Antennas}, Hamburg, Germany, February 18-20, 2020.

\bibitem{ref13} M. Simunek, F. P. Font\'{a}n and P. Pechac, "The UAV Low Elevation Propagation Channel in Urban Areas: Statistical Analysis and Time-Series Generator," \textit{IEEE Trans. on Ant. and Propagat}., vol. 61, no. 7, pp. 3850-3858, July 2013, DOI: \href{https://doi.org/10.1109/TAP.2013.2256098}{10.1109/TAP.2013.2256098}.

\bibitem{ref14} J. Zeleny, F. P\'{e}rez-Font\`{a}n and P. Pechac, "Initial Results from a measurement campaign for low elevation angle links in different environments," \textit{9th European Conference on Antennas and Propagation} (EuCAP), Lisbon, Portugal, April 12-17, 2015.

\bibitem{ref15} D. W. Matolak and R. Sun, ``Air\textendash{}ground channel characterization for unmanned aircraft systems: The near-urban environment,'' \textit{Proc. IEEE Mil. Comm. Conf. (MILCOM)}, Tampa, FL, USA, October 26-28, 2015.

\bibitem{ref16} M. Bucur, T. Sorensen, R. Amorim, M. Lopez, I. Z. Kovacs and P. Mogensen, "Validation of Large-Scale Propagation Characteristics for UAVs within Urban Environment," \textit{IEEE 90th Veh. Tech. Conf. (VTC2019-Fall)}, Honolulu, HI, USA, September 22-25, 2019.

\bibitem{ref17} M. Lopez, T. B. Sorensen, P. Mogensen, J. Wigard and I. Z. Kovacs, "Shadow Fading Spatial Correlation Analysis for Aerial Vehicles: Ray Tracing vs. Measurements\textit{," IEEE 90th Veh. Tech. Conf.} (VTC2019-Fall), Honolulu, HI, USA, September 22-25, 2019.

\bibitem{ref18} S. Ranvier, J. Kivinen and P. Vainikainen, "Millimeter-Wave MIMO Radio Channel Sounder", IEEE Trans. on Instr. and Meas., vol. 56, no. 3, June 2007, pp. 1018-1024, DOI: \href{https://doi.org/10.1109/TIM.2007.894197}{10.1109/TIM.2007.894197}.

\bibitem{ref19} V-M. Kolmonen, P. Almers, J. Salmi, J. Koivunen, K. Haneda, A. Richter, F. Tufvesson, A.F. Molisch, P. Vainikainen, "A Dynamic Dual-Link Wideband MIMO Channel Sounder for 5.3 GHz", \textit{IEEE Trans. on Instr. and Meas.}, vol. 59, no. 4, April 2010, pp. 873-883, DOI: \href{https://doi.org/10.1109/TIM.2009.2026608}{10.1109/TIM.2009.2026608}.

\bibitem{ref20} A. Mohammed. Al-S, Tharek A. Rahman, M. H. Azmi, S.A. Al-Gailani, ``Millimeter-wave propagation measurements and models at 28\,GHz and 38\,GHz in a dining room for 5G wireless networks'', \textit{Measurement}, vol. 130, December 2018, pp. 71-81, DOI: \href{https://doi.org/10.1016/j.measurement.2018.07.073}{10.1016/j.measurement.2018.07.073}.

\bibitem{ref21} W. G. Newhall et al., "Wideband air-to-ground radio channel measurements using an antenna array at 2 GHz for low-altitude operations," \textit{Proc.} \textit{IEEE Mil. Comm. Conf. (MILCOM)}, Boston, MA, USA, October 1-6, 2003.

\bibitem{ref22} D. W. Matolak and R. Sun, ``Antenna and frequency diversity in the unmanned aircraft systems bands for the over-sea setting,'' in \textit{Proc. IEEE Digit. Avionics Syst. Conf. (DASC)}, Colorado Springs, CO, USA, October 5-9, 2014.

\bibitem{ref23} J. Chen, B. Daneshrad, and W. Zhu, ``MIMO performance evaluation for airborne wireless communication systems,'' \textit{Proc. Mil.} \textit{Comm. Conf. (MILCOM)}, Baltimore, MD, USA, November 7-10, 2011.

\bibitem{ref24} Aslatech (\url{https://it.linkedin.com/in/asla-aslatech/})

\bibitem{ref25} PX4 Firmware (\url{https://github.com/PX4/PX4-Autopilot/releases/tag/v1.9.0/})

\bibitem{ref26} \url{https://spectrumcompact.com/}

\bibitem{ref27} \url{http://qgroundcontrol.com/}

\bibitem{ref28} \url{https://mavlink.io/en/}

\bibitem{ref29} Ente Nazionale per l’Aviazione Civile (ENAC): \url{https://www.enac.gov.it/en}

\bibitem{ref30} European Union Aviation Safety Agency (EASA): \url{https://www.easa.europa.eu/domains/civil-drones-rpas}

\bibitem{ref31} V. Degli-Esposti, F. Fuschini, and D. Guiducci, ``A study on roof-to-street propagation,'' in \textit{2003 ICEAA Proc. Of International Conference on Electromagnetics in Advanced Applications}, 2003, pp. 45\textendash{}47.

\bibitem{ref32} J. Walfisch, H. L. Bertoni, ``A Theoretical Model of UHF Propagation in Urban Environments,'' \textit{IEEE Trans. on Ant. and Propagat.}, Vol. 36 No. 12, pp. 1788 \textendash{} 1796.

\bibitem{ref33} E. M. Vitucci, V. Semkin, M. J. Arpaio, M. Barbiroli, F. Fuschini, C. Oestges, V. Degli-Esposti, ``Experimental Characterization of Air-to-ground Propagation at mm-Wave Frequencies in Dense Urban Environment,'' in Proc. of 15$^{\mathrm{th}}$ European Conference on Antennas and Propagation (EuCAP 2021), D\"{u}sseldorf, Germany, 22-26 March 2021.

\bibitem{ref34} V. Degli-Esposti, F. Fuschini, E.M. Vitucci, G. Falciasecca, "Measurement and modelling of scattering from buildings," in \textit{IEEE Trans. on Ant. and Propagat.}, vol. 55, no. 1, pp. 143-153, Jan 2007, DOI: \href{https://doi.org/10.1109/TAP.2006.888422}{10.1109/TAP.2006.888422}.

\bibitem{ref35} V. Degli-Esposti, J.S. Lu, J.N. Wu, J.J. Zhu, J.A. Blaha, E.M. Vitucci, F. Fuschini, M. Barbiroli, ``A Semi-Deterministic Model for Outdoor-to-Indoor Prediction in Urban Areas, \textit{IEEE Ant. and Wirel. Propagat. Letters}, vol. 16, June 2017, pp. 2412-2415, DOI: \href{https://doi.org/10.1109/LAWP.2017.2721739}{10.1109/LAWP.2017.2721739}.

\bibitem{ref36} V. Semkin, E. M. Vitucci, F. Fuschini, M. Barbiroli, V. Degli-Esposti, and C. Oestges, ``Characterizing The Drone-to-Machine UWB Radio Channel in Smart Factories,'' submitted to \textit{IEEE Access}, 2021.

\end{thebibliography}

\end{document}